\pdfoutput=1
\documentclass[aps,prx,reprint,showpacs,superscriptaddress,notitlepage,floatfix]{revtex4-2}

\usepackage[pdftex]{graphicx}
\usepackage[pdftex]{epsfig}
\usepackage{epstopdf}
\usepackage{color}
\usepackage{amssymb,amsmath}
\usepackage{graphicx}
\usepackage{dcolumn}
\usepackage{multirow}
\usepackage{hyperref}
\usepackage{siunitx}
\usepackage{soul}

\usepackage{mathtools}

\usepackage[normalem]{ulem}

\usepackage{enumitem,amssymb}
\newlist{todolist}{itemize}{2}
\setlist[todolist]{label=$\square$}

\begin{document}

\title{Hole in one: Pathways to deterministic single-acceptor incorporation in Si(100)-2$\times$1}

\author{Quinn Campbell}
\affiliation{Center for Computing Research, Sandia National Laboratories, Albuquerque NM, USA}
\email[Corresponding Author: ]{qcampbe@sandia.gov}
\author{Andrew D. Baczewski}
\affiliation{Center for Computing Research, Sandia National Laboratories, Albuquerque NM, USA}
\author{R. E. Butera}
\affiliation{Laboratory for Physical Sciences, College Park MD, USA}
\author{Shashank Misra}
\affiliation{Sandia National Laboratories, Albuquerque NM, USA}

\begin{abstract}
Stochastic incorporation kinetics can be a limiting factor in the scalability of semiconductor fabrication technologies using atomic-precision techniques.
While these technologies have recently been extended from donors to acceptors, the extent to which kinetics will impact single-acceptor incorporation has yet to be assessed.
We develop and apply an atomistic model for the single-acceptor incorporation rates of several recently demonstrated precursor molecules: diborane (B$_2$H$_6$), boron trichloride (BCl$_3$), and aluminum trichloride in both monomer (AlCl$_3$) and dimer forms (Al$_2$Cl$_6$), to identify the acceptor precursor and dosing conditions most likely to yield deterministic incorporation.
While all three precursors can achieve single-acceptor incorporation, we predict that diborane is unlikely to achieve deterministic incorporation, boron trichloride can achieve deterministic incorporation with modest heating (\SI{50}{\celsius}), and aluminum trichloride can achieve deterministic incorporation at room temperature. 
We conclude that both boron and aluminum trichloride are promising precursors for atomic-precision single-acceptor applications, with the potential to enable the reliable production of large arrays of single-atom quantum devices.
\end{abstract}

\maketitle
\section*{Introduction}
Single shallow dopants placed with atomic precision in silicon might be used to realize qubits~\cite{o2001towards,buch2013spin,hill2015surface,watson2017atomically,hile2018addressable,he2019two,keith2019single,kranz2020exploiting,laucht2021roadmap,bussmann2021atomic}, single-to-few-carrier devices~\cite{fuechsle2010spectroscopy,fuechsle2012single,koch2019spin,wyrick2019devices,anderson2020low,wang2020atomic,ward2020atomic}, and analog quantum simulators~\cite{salfi2016quantum,le2017extended,dusko2018adequacy,le2020topological,altman2021quantum}.
While the placement of phosphorus donors using a phosphine (PH$_3$) precursor has received the most development~\cite{schofield2003placement,ruess2004toward,Wilson2004dissociation}, recent demonstrations involving arsenic~\cite{stock2020atomic}, boron~\cite{vskerevn2020bipolar,dwyer2021area}, and aluminum~\cite{radue2021alcl3,owen2021alkyls} indicate that the breadth of viable chemistries is growing.
The acceptors, in particular, offer opportunities that are complementary to donors due to their relatively large spin-orbit coupling~\cite{ruskov2013chip,salfi2016quantum1,salfi2016charge}, the absence of valley-orbit coupling~\cite{koiller2001exchange,salfi2014spatially,salfi2018valley,voisin2020valley}, and suppressed hyperfine interaction~\cite{stegner2010isotope,philippopoulos2019hole}.


These features have led to significant interest in single acceptors in silicon~\cite{calvet2007effect,calvet2007observation,mol2013interplay,van2014probing,abadillo2015interface,abadillo2018entanglement,van2018readout}, notably finding that with appropriate strain manipulation, long coherence times of $\sim$10 ms can be achieved~\cite{kobayashi2020engineering}.
Arrays of such acceptors present possibilities for analog quantum simulation of the extended Fermi--Hubbard model~\cite{salfi2016quantum}, which has motivated several theoretical investigations~\cite{durst2020quadrupolar,zhu2020linear,zhu2021multi}.
We also note the prospect of ultra doping silicon with acceptors to realize superconductivity, and with that the possibility of fabricating a variety of quantum devices (e.g., wires, Josephson junctions, SQUIDs, etc.)~\cite{bustarret2006superconductivity,shim2014bottom,duvauchelle2015silicon,bonnet2021strongly}.
While these applications motivate further research into the possibility of single-acceptor quantum devices, until recently no acceptor precursors were known to be compatible with atomic-precision fabrication techniques, let alone whether they can be used to achieve deterministic incorporation.

\begin{figure}[t]
    \includegraphics[width=\columnwidth]{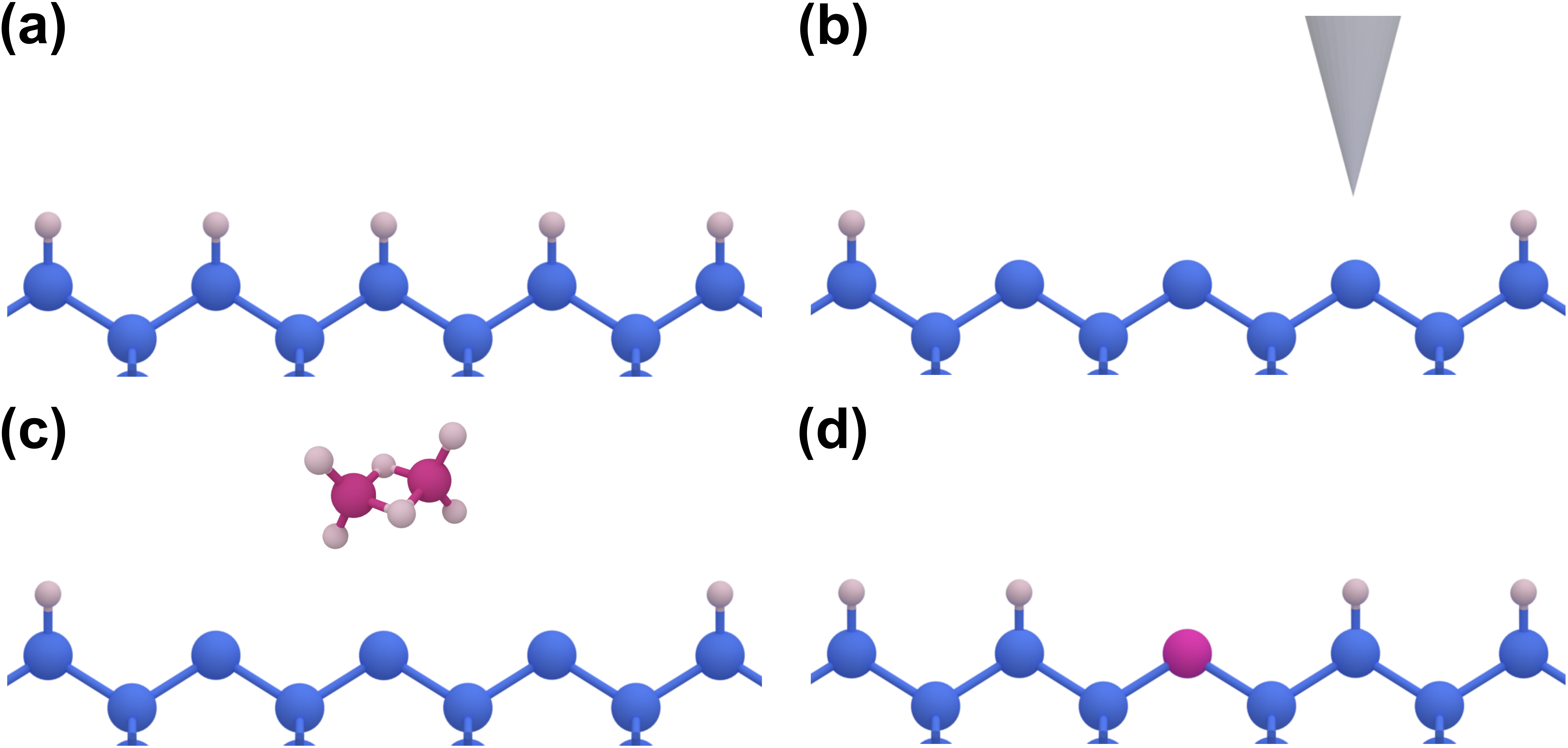}
    \caption{
    A schematic overview of the process for atomic-precision placement of dopants modeled throughout this paper. A silicon(100)-2$\times$1 surface is prepared and hydrogen terminated (a). A small window, typically three silicon dimers wide, is then depassivated using a lithography tool such as scanning tunneling microscopy (STM) (b). This window is then exposed to a precursor gas (for example, in this figure, diborane) (c) which preferentially adsorbs on the bare silicon and not the surrounding hydrogen resist. This precursor molecule then dissociates within the exposed window, eventually incorporating its acceptor atom into the surface (d).
    }
    \label{fig:overview}
\end{figure}

Atomic-precision placement of dopants in silicon has been achieved for phosphorus donors by first preparing a hydrogen- (or halogen-~\cite{pavlova2018first,dwyer2019stm,silva2020reaction,pavlova2021reactivity}) terminated Si(100)-2x1 surface, as shown in Fig.~\ref{fig:overview}.
A lithography tool such as a scanning tunneling microscope (STM) is then used to remove a small portion of the resist~\cite{lyding1994nanoscale,shen1995atomic,randall2009atomic}.
To introduce a single dopant with a three-fold coordinated precursor, this pattern should be at least two or three dimers long in the dimer row direction to make room for the removal of the coordinating atoms. 
Under ultrahigh vacuum, the surface is then exposed to a pure, low-pressure gas-phase source of a precursor molecule which selectively adsorbs onto the bare silicon, but not the surrounding resist.
This precursor molecule will then dissociate on the bare silicon surface, leading to the eventual incorporation of a dopant atom. 
For a sufficiently selective atomic resist, this will be confined to the initial depassivation window, leading to atomic-precision placement of the dopant i.e., within $\pm$ 1 lattice sites of the target~\cite{schofield2003placement}.
This methodology has recently been extended to acceptor $\delta$-doping layers of silicon using precursors of diborane (B$_2$H$_6$), boron trichloride (BCl$_3$) and aluminum trichloride (AlCl$_3$)~\cite{vskerevn2020bipolar,radue2021alcl3,dwyer2021area}, but these techniques have not yet been demonstrated for a single acceptor.

Furthermore, to reliably fabricate devices consisting of many precisely placed single acceptor atoms (e.g., the sites in an analog quantum simulator for the Fermi-Hubbard model) the probability of incorporation needs to be deterministic, i.e. $\approx$1.
While the dissociation and incorporation pathways for most precursor molecules are thermodynamically downhill and thus highly likely to happen, the exact geometry and speed of dissociation depends on the initial dosing conditions such as precursor pressure, exposure time, and sample temperature. 
Assuming that the probabilities of incorporation are independent from site to site, the probability of successfully fabricating an array with no missing sites will decay exponentially in the total number of sites.
For example, even in using a precursor and process with a 90\% probability of single-dopant incorporation for each site, the probability of creating a 3$\times $3 array without missing sites is only $\approx$40\%~\footnote{Assuming that the probability of writing a specific lithographic window is 1 and that the failure to incorporate due to kinetic effects is independent from site to site, this statistic is derived as $(0.9)^{3\times 3}\approx 0.4$.}!
In other words, the process of a single acceptor incorporating with atomic-precision must be deterministic, not stochastic, to be realistically scalable.
Previous work on phosphine has estimated this single-donor incorporation percentage at $\approx$ 65\% when dosed at room temperature, although corroborating kinetic modeling revealed a potential pathway toward deterministic incorporation by increasing the dosing temperature to $\sim$ \SI{150}{\celsius}~\cite{ivie2021impact}.

In this work, we develop several Kinetic Monte Carlo (KMC) models based on previously calculated first principles reaction pathways~\cite{campbell2021model,dwyer2021impact,radue2021alcl3} to predict the single-acceptor incorporation statistics of three recently explored atomic-precision acceptor precursors~\cite{vskerevn2020bipolar,radue2021alcl3,dwyer2021area}: diborane, boron trichloride, and aluminum trichloride in both monomer and dimer forms.
We demonstrate that while single-acceptor atomic-precision incorporation is possible for all precursors, they are not equally likely to incorporate.
We predict that diborane will likely not be able to realize deterministic incorporation, boron trichloride realizes deterministic incorporation with modest heating, and aluminum trichloride realizes deterministic incorporation at room temperature in both dimer and monomer forms.
Our work implies boron and aluminum trichloride can be reliably used to create arrays of single acceptors placed with atomic precision on Si(100)-2x1, potentially enabling the large-scale production of single-atom transistors, qubits, and analog quantum simulation devices.

\section*{Results}
\subsection*{Diborane (B$_2$H$_6$)}

\begin{figure*}
    \includegraphics[width=\textwidth]{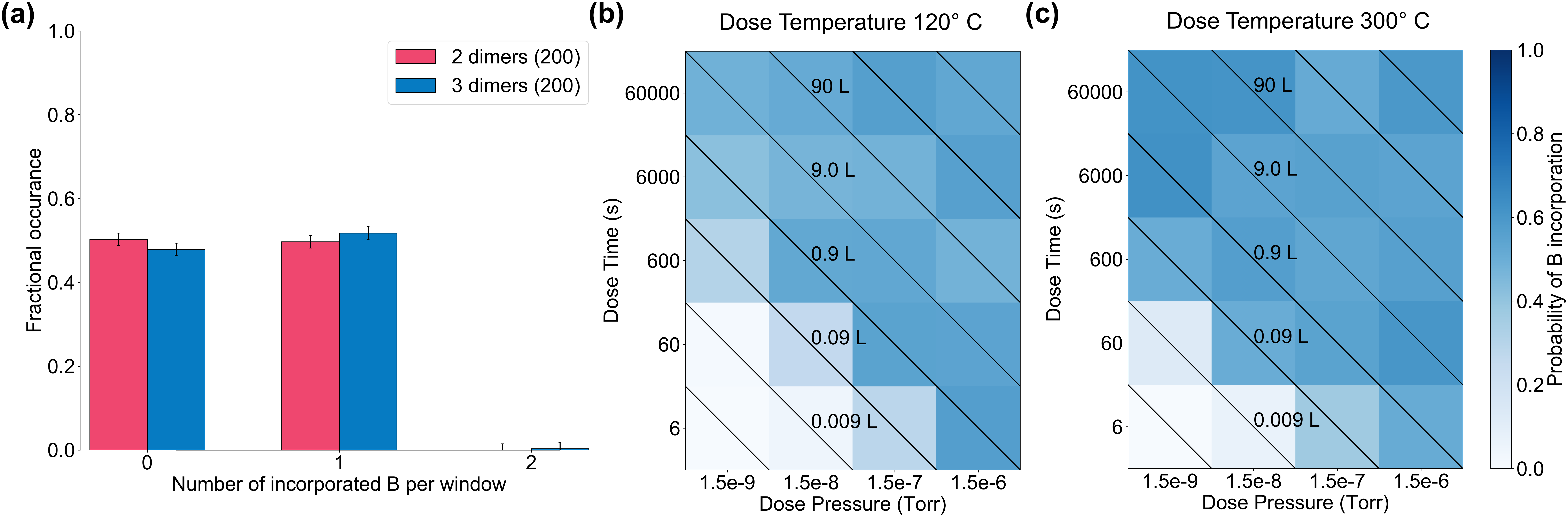}
    \caption{
    (a) The probability of B incorporation using a diborane precursor with a dosing pressure of 1.5 $\times$ 10$^{-7}$ Torr for 10 minutes at \SI{120}{\celsius}, followed by an anneal at \SI{410}{\celsius} for 1 minute. (b) The probability of incorporation using a diborane precursor as a function of dose pressure and exposure time for a dosing temperature of \SI{120}{\celsius}. (c) The probability of incorporation using a diborane precursor as a function of dose pressure and exposure time for a dosing temperature of \SI{300}{\celsius}.
    }
    \label{fig:diborane}
\end{figure*}

We begin by considering a KMC model for diborane, the first acceptor precursor to be successfully implemented using an APAM-like process for $\delta$-doping~\cite{vskerevn2020bipolar}.
Diborane requires significant heating during dosing to increase its sticking coefficient on bare silicon, and then a high anneal temperature (~$\sim$ \SI{400}{\celsius} to achieve significant levels of incorporation).
Campbell \textit{et al.} subsequently rationalized that this was due to a complex dissociation pathway with high reaction barriers and the need to break apart the dimer precursor~\cite{campbell2021model}.
In this paper, we use the same set of reactions to predict the single-acceptor incorporation rates for windows two and three silicon dimers wide.
Following the experimental setup of {\v{S}}kere{\v{n}}~\textit{et al.}, we use a dosing pressure of 1.5 $\times$ 10$^{-7}$ Torr for 10 minutes at \SI{120}{\celsius}, followed by an anneal at \SI{410}{\celsius} for 1 minute for our initial simulation, shown in Fig.~\ref{fig:diborane}a.
At these dosing conditions, diborane shows a particularly low incorporation rate, only reaching 52\% $\pm$ 1.8 \% for a three silicon dimer wide window.
This low incorporation rate can be attributed to the relative complexity of the dissociation pathway, which requires overcoming a barrier of at least 1.3 eV for an incorporation event.
Furthermore, at these dosing conditions, the dissociation of the diborane is often impeded by the adsorption of another diborane within the dimer window, occupying valuable space for either molecule to shed its hydrogen and split its dimer bond.

To test whether the predicted low rates of single-acceptor incorporation are  an artifact of these particular dosing settings, we calculate the incorporation rate across a broad array of dosing pressures and exposure times, with results illustrated in Fig.~\ref{fig:diborane}b.
We do not predict high rates of incorporation at any of the considered pressures and times, with the highest probability of incorporation in a three silicon dimer window at 57\% $\pm$ 1.7 \% occurring at a dose pressure of \SI{1.5e-6}{Torr} and an exposure time of 6 seconds.
All of these calculations, however, are at a temperature of \SI{120}{\celsius}.
Increasing the temperature at which the dosing occurs would increase the likelihood of any given diborane molecule fully dissociating before another adsorbs.
This will also increase the likelihood of overcoming the necessary $\gtrsim$ 1.3 eV barriers for incorporation.
However, our model predicts that hot dosing the sample at \SI{300}{\celsius} (close to typical anneal temperatures), as shown in Fig.~\ref{fig:diborane}c, results in an insignificant increase in incorporation rates.
At \SI{300}{\celsius}, the highest rate of incorporation achieved is 63\% $\pm$ 1.6\%, at a dose pressure of \SI{1.5e-9}{Torr} and an exposure time of 6000 seconds ($\sim$1.7 hours).
Even if incorporation were likely at this temperature, loss of pattern fidelity would likely lead to poor precision in location anyway.
We thus conclude that while diborane is useful for $\delta$-doping, it is unlikely to be a particularly promising precursor for single-acceptor applications.

Our kinetic model requires reaction barriers as inputs which have been generated by Campbell \textit{et al.} using Density Functional Theory (DFT)~\cite{campbell2021model}.
DFT is well known to have errors of order $\sim$0.1 eV in adsorption energies and reaction barriers. 
Given that these reaction barriers appear in the exponential of the Arrhenius rates in our kinetic model, it is imperative that we test the sensitivity of our results to standard DFT errors.
This is particularly pronounced in diborane: one of the critical decision points for whether diborane will incorporate is whether its boron dimer will initially break up after adsorption or the molecule will shed hydrogen to nearby silicon dimers.
Both of these reactions have almost the same barrier, 0.91 eV and 0.89 eV, making the decision as to which reaction path will be followed essentially a coin toss. 
By lowering one of these barriers by 0.1 eV, we can essentially control which dissociation pathway is followed, which leads to drastically increased or decreased levels of incorporation.

By lowering the reaction barrier for the boron dimer splitting up after initial adsorption by 0.1 eV, the probability of getting a single incorporation within a three dimer window is increased to 94\% with a dose pressure of 1.5 $\times$ 10$^{-7}$ Torr for 10 minutes at \SI{120}{\celsius}. 
In contrast, lowering the reaction barrier for the adsorbed diborane to instead shed hydrogen and keep its boron dimer intact by 0.1 eV reduces the single-acceptor incorporation rate to a mere  5\%.
Manipulating other reactions within the diborane pathway results in negligible changes to the single-acceptor incorporation rate.
This demonstrates that the key factor in testing the sensitivity of a barrier to reasonable levels of error is whether this barrier changes what dissociation pathway is most likely to occur. 
Diborane, in this case, is uniquely sensitive to errors in the reaction barriers right after initial adsorption due to the importance in determining the overall dissociation path.
Future work should either focus on using higher accuracy techniques to provide the reaction barriers for this initial dimer dissociation decision point or otherwise experimentally resolve this point.

\subsection*{Boron trichloride (BCl$_3$)}

\begin{figure*}[ht]
    \includegraphics[width=\textwidth]{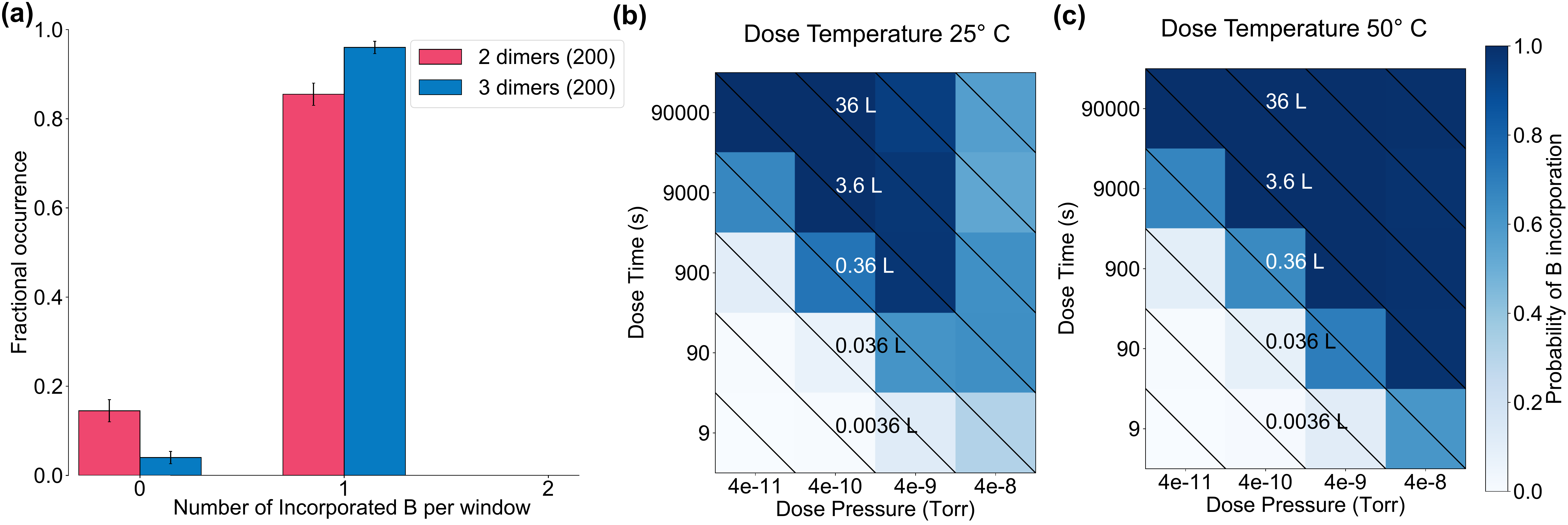}
    \caption{
    (a) The probability of B incorporation using a BCl$_3$ precursor with a dosing pressure of \SI{4e-9}{Torr}, an exposure time of 900 seconds, and a dose temperature of \SI{25}{\celsius}. (b) The probability of incorporation using a BCl$_3$ precursor as a function of dose pressure and exposure time for a dosing temperature of \SI{25}{\celsius}. (c) The probability of incorporation using a BCl$_3$ precursor as a function of dose pressure and exposure time for a dosing temperature of \SI{50}{\celsius}. Deterministic single-acceptor incorporation is achieved for all doses $>$ 1 L. 
    }
    \label{fig:bcl3}
\end{figure*}

We next turn to boron trichloride (BCl$_3$), which has recently been demonstrated to have excellent acceptor doping properties in $\delta$-doped layers~\cite{dwyer2021area}.
In contrast to diborane, boron trichloride exhibits a drastically simplified reaction pathway as demonstrated by Dwyer \textit{et al.}\cite{dwyer2021impact}, with only three possible steps and a likely reaction barrier of 0.93 eV. 
This results in greatly increased levels of incorporation, even at room temperature, as demonstrated in Fig.~\ref{fig:bcl3}a.
Matching the experimental conditions of Dwyer \textit{et al.} with a dosing pressure of \SI{4e-9}{Torr}, an exposure time of 900 seconds, and a dose temperature of \SI{25}{\celsius}, BCl$_3$ gives a probability of a single B atom incorporating of 96\% $\pm$ 0.3\% within a three silicon dimer wide window.
This percentage is still relatively high even with only two silicon dimers to dissociate within: 86\% $\pm$ 0.9\%.

The incorporation rate can be increased to near-deterministic levels by altering the dosing conditions.
As seen in Fig.~\ref{fig:bcl3}b, combinations of low pressure and temperature, such as \SI{4e-10}{Torr} of BCl$_3$ exposed for 9000 seconds (2.5 hours) can achieve deterministic single-acceptor incorporation.
Given the simplistic nature of the BCl$_3$ dissociation pathway, we can easily identify two mechanisms that will limit deterministic incorporation -- either no molecule adsorbs in the window at low pressure or too many molecules adsorb at high pressure.
If these conditions are overcome, then deterministic incorporation will occur.
First, at least one BCl$_3$ molecule must adsorb for any dissociation to happen.
This explains the low incorporation probabilities seen at low pressures and short exposure times and even at levels of fractional Langmuir dose.
While a dose of 0.36 L should guarantee at least one adsorption within a three silicon dimer window, there are still rare occasions ($\approx$ 4\%) where no adsorption occurs within the dosing window.
Moving to higher dose levels ensures that there will always be some level of dosing, explaining the move toward deterministic incorporation at \SI{4e-10}{Torr} by increasing the dose time from 900 to 9000 seconds.
The second mechanism for sub-deterministic incorporation, which applies at higher pressures, is not a deficit of adsorbing BCl$_3$ molecules, but instead a surplus.
The BCl$_3$ molecule needs to have sufficient time and space to dissociate and this can only be easily achieved if there are no additional BCl$_3$ molecules in the three silicon dimer window, blocking further dissociation.
At higher pressures, additional BCl$_3$ molecules are likely to adsorb into the depasivated  window before the first BCl$_3$ molecule has had time to fully dissociate, leading to no incorporation events.

Deterministic incorporation of a single boron atom from a BCl$_3$ precursor can therefore be achieved by ensuring that (1) a single BCl$_3$ molecule is adsorbed and (2) the first BCl$_3$ molecule has enough time and space to dissociate without competition from other BCl$_3$ molecules. 
Addressing (1) is easy enough: simply choose doses $>$ 1 L. 
Addressing (2) requires a bit more finesse. 
At room temperature, (2) can be achieved by dosing at low pressures for long times, which gives BCl$_3$ enough time to dissociate before the next likely adsorption event. 
These low pressure and long exposure times scenarios are cumbersome at best, however, and completely unrealistic at worst. 
At these low pressures and long exposure times, the background fractions of water and other species will be a significant fraction of a dose, leading to an increase in contamination within a patterned region. 
An alternate method is to increase the speed at which BCl$_3$ dissociation takes place by heating the system during dosing. 
As shown in Fig.~\ref{fig:bcl3}c, even minor levels of heating can achieve the desired effect, with all doses $>$ 1 L displaying deterministic incorporation with heating to only \SI{50}{\celsius}.

Boron trichloride is much less sensitive than diborane to potential errors in the reaction barriers of order 0.1 eV. 
Due to the simplified reaction pathway, increasing the main barrier for BCl$_3$ dissociation by 0.1 eV merely increases the temperature at which deterministic incorporation can be achieved at all $>$ 1 L doses to \SI{75}{\celsius}.
Similarly, lowering this barrier by 0.1 eV results in deterministic single-acceptor incorporation at room temperature at all $>$ 1 L doses. 
We conclude that, with only minor levels of heating, BCl$_3$ is a promising precursor for applications that require deterministic single-acceptor placement with atomic precision.

\subsection*{Aluminum trichloride (AlCl$_3$)}
\begin{figure*}
    \includegraphics[width=\textwidth]{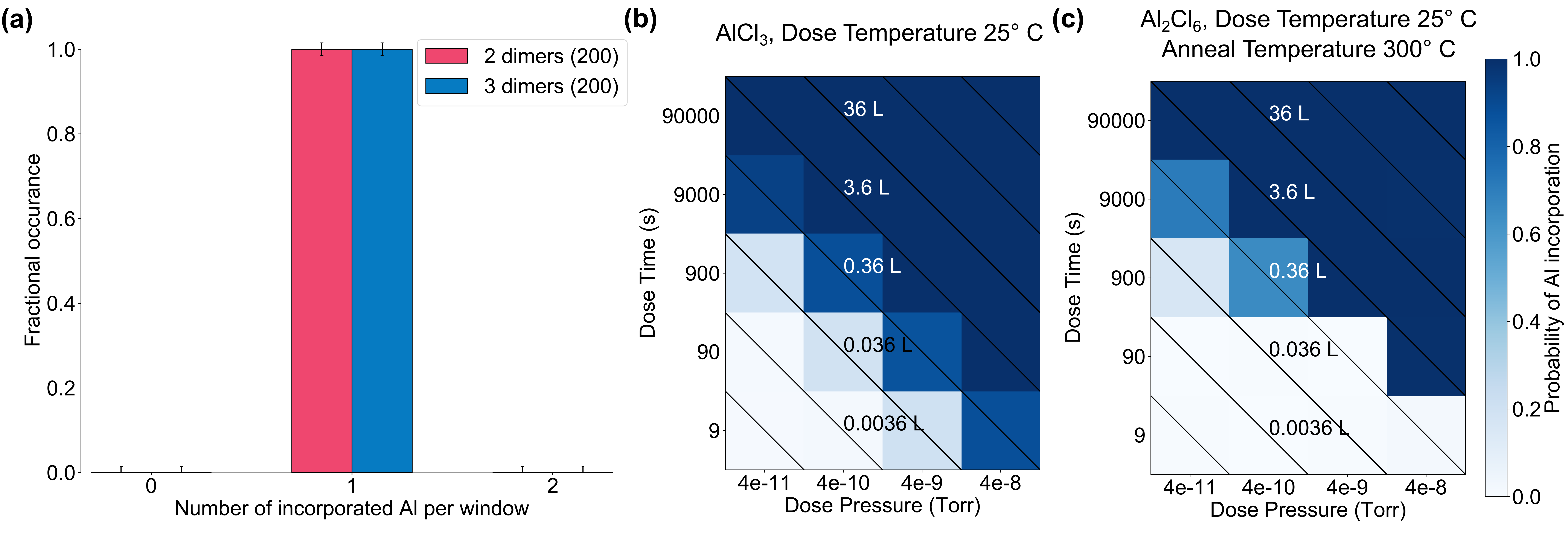}
    \caption{
     (a) The probability of Al incorporation using a AlCl$_3$ precursor with a dosing pressure of \SI{4e-9}{Torr}, an exposure time of 900 seconds, and a dose temperature of \SI{25}{\celsius}. (b) The probability of incorporation using a AlCl$_3$ monomer precursor as a function of dose pressure and exposure time for a dosing temperature of \SI{25}{\celsius}. (c) The probability of incorporation using a aluminum trichloride dimer (Al$_2$Cl$_6$) precursor as a function of dose pressure and exposure time for a dosing temperature of \SI{25}{\celsius}. Deterministic single-acceptor incorporation is achieved for all doses $>$ 1 L in both the monomer and dimer case. 
    }
    \label{fig:alcl3}
\end{figure*}
Finally, we turn our KMC model to aluminum trichloride, which has been recently demonstrated as a viable acceptor precursor by Radue \textit{et al.}~\cite{radue2021alcl3}.
We predict AlCl$_3$ to be an the most promising precursor for single-acceptor applications.
As shown in Fig.~\ref{fig:alcl3}a, with a dosing pressure of \SI{4e-9}{Torr} and an exposure time of 900 seconds, deterministic incorporation is achieved in both two- and three-dimer windows at room temperature!
In Fig.~\ref{fig:alcl3}b, we demonstrate that this result is robust to dosing conditions. 
The AlCl$_3$ monomer will achieve deterministic single-acceptor incorporation at any dose $>$ 1 L at room temperature. 
These settings are easily accessible in an experimental setting and provide the most potential for deterministic incorporation without additional heating of any of the precursors studied.
Why does AlCl$_3$ exhibit superior incorporation to BCl$_3$? 
In comparison to BCl$_3$, the pathway for AlCl$_3$ dissociation remains exactly the same.
The reaction barriers, however, are lowered by $\sim$ 0.3 eV. 
This makes the dissociation pathway much easier to achieve in a timely manner without the need for additional heating.

The picture for aluminum trichloride is complicated, however, by the fact that in gas form it often appears as both a dimer as well as a monomer~\cite{tomita1983high}. 
It is possible, and perhaps even likely, that the effort of breaking apart an initial dimer will overwhelm the advantage of low barriers that the AlCl$_3$ monomer enjoys. 
In Fig.~\ref{fig:alcl3}c, we show the incorporation statistics for an AlCl$_3$ dimer (from here on referred to as Al$_2$Cl$_6$) with a dose at room temperature and a subsequent anneal of \SI{350}{\celsius}.
We again see deterministic incorporation at all doses $>$ 1 L.
The subsequent anneal, however, is now crucial to achieving deterministic incorporation: with a dosing pressure of \SI{4e-10}{Torr} and an exposure time of 900 seconds but no high temperature anneal, Al$_2$Cl$_6$ only sees a 54\% $\pm$ 1.8\% probability of incorporation.
This is in contrast to BCl$_3$ and AlCl$_3$ which need no anneal to reach deterministic incorporation within our model. 
The need for an anneal to reach substantial levels of incorporation in Al$_2$Cl$_6$ can be explained by the higher barrier of 1.26 eV needed to split the dimer into two separate AlCl$_x$ fragments. 
Once the higher temperature anneal has split the dimer, it is then likely to dissociate quickly in the same manner as the AlCl$_3$ monomer. 

Also crucial to the deterministic outcome is the balance of reaction barriers for splitting the aluminum dimer versus shedding chlorine. 
In diborane we saw that these reaction barriers were almost exactly the same, leading to difficulty in achieving the preferred outcome (dimer splitting) for eventual incorporation.
In Al$_2$Cl$_6$, the reaction barrier for breaking up the aluminum dimer is 1.26 eV, while the reaction barrier for shedding chlorine is 1.49 eV.
This is a large enough difference in reaction barriers to ensure that the just adsorbed Al$_2$Cl$_6$ will almost always break apart into separate AlCl$_{\rm x}$ fragments, which can then easily dissociate into incorporating positions. 

Another factor complicating the incorporation statistics of Al$_2$Cl$_6$ is the possibility of two incorporated aluminum atoms dimerizing and becoming electrically inactive. 
Prior work on boron clustering in silicon has shown electrically inactive dimer complexes may rationalize the presence of immobile boron in studies on ion implanted samples~\cite{tarnow1992theory,stolk1995implantation,zhu1996ab}, and it is reasonable to believe there might be a similar effect for aluminum.
In our model we consider sequences of reactions that lead to an AlCl bridging two different silicon dimers as an incorporation event because the probability of desorbing from this state is vanishingly small (see the Methods section  and Fig.~\ref{fig:endpoints} for more details).
We assign two bridging AlCl fragments occurring in the same stretch of silicon as a dimerized configuration which will be inactive and do not count it as an incorporation event within our model. 
Nonetheless, we see deterministic incorporation of a single Al atom because the three silicon dimer wide window is only wide enough to support the dissociation of one of the AlCl$_2$ fragments.
Desorption of two Cl atoms to form Cl$_2$(g) could potentially free up the necessary space, but it does not occur within a reasonable timeframe during a \SI{350}{\celsius} anneal.
The final configuration after dissociation of Al$_2$Cl$_6$ is therefore always a bridging AlCl fragment next to a AlCl$_2$ fragment, which our model treats as a single incorporation.
It is worth examining this assignment of outcomes, however, as it implicitly assumes that the AlCl$_2$ fragment will either desorb or diffuse away during subsequent processing steps, which is not inherently clear from physical experiments. 
If we instead assume that a bridging AlCl fragment next to a AlCl$_2$ fragment will eventually dimerize and become electrically inactive, the predicted incorporation rate falls to zero for all dosing conditions examined!
We thus conclude that AlCl$_3$ as a monomer is an ideal precursor for deterministic atomic-precision single-acceptor incorporation, and that the dimer Al$_2$Cl$_6$ is likely also an exceptional candidate, but further work is needed to elucidate the exact dimerization mechanisms during subsequent processing steps.

Aluminum trichloride, similar to boron trichloride, exhibits little sensitivity to errors in calculated barriers. 
In its the monomer form, an increase of the main dissociation barrier by 0.1 eV still ends with deterministic incorporation at room temperature, as the barrier remains low enough to be easily overcome without additional heating. 
In dimer form, the deterministic incorporation rate is also robust to increasing the barrier of a dimer splitting apart by 0.1 eV. 
This is because, in the dimer case, all the dissociation takes place in the anneal step at  \SI{350}{\celsius} which is also sufficient to overcome the newly raised barrier. 

\section*{Discussion}
\label{sec:discussion}

We have used kinetic models based on first principles calculations of dissociation barriers to calculate the single-acceptor incorporation statistics for diborane, boron trichloride, and aluminum trichloride, in both monomer and dimer form. 
We demonstrated that while diborane exhibits poor incorporation statistics, a coin flip at best, both boron and aluminum trichloride can achieve deterministic single-acceptor incorporation with no (or minimal) heating beyond room temperature. 
This suggests that boron and aluminum trichloride are viable precursors to create large-scale atomic-precision arrays of single acceptors with extremely low defect rates.
Overall, the methods demonstrated in this work present potential pathways for the large-scale production of qubits, single-to-few-carrier devices, and analog quantum simulators.

It is likely that there are mitigation strategies for precursors/dosing conditions with stochastic incorporation, with certain trade offs.
For example, targeting devices comprised of an uncertain number of dopants greater than one and designing around variability in device parameters.
The work presented here implies that using larger windows for boron and aluminum trichloride at reasonable conditions will likely yield at least one acceptor incorporation (and probably more), allowing greater room for error in processing.
Further study on interactions of these precursors across different silicon dimer rows would be necessary to develop a comparable kinetic model.

One factor that may complicate the performance of single-acceptor devices made from chloride precursors is the shedding of chlorine to nearby silicon.
While recent work has shown that chloride precursors are compatible with a hydrogen resist~\cite{dwyer2021area}, it is not clear that residual chlorine atoms will not interfere with the incorporation of electrically active acceptors.
This problem might be overcome by substantial thermal anneals or even UV photolithography to remove the extraneous atoms~\cite{katzenmeyer2021photothermal}, though this may have deleterious impacts on the precision of the placement.

Additionally, acceptor doping with boron presents additional complications due to isotopic purity not typically encountered when doping with phosphorus due to the existence of two stable isotopes with distinct nuclear spins: $^{10}$B (spin-3, natural abundance $\approx$20\%) and $^{11}$B (spin-3/2, natural abundance $\approx$80\%)~\cite{meija2016isotopic}.
In contrast, $^{27}$Al is the only stable isotope of aluminum (spin-5/2)~\cite{meija2016isotopic}.
It is worth noting that all of the stable isotopes of both species have quadrupolar nuclei and are thus amenable to coherent manipulation through the application of strain, magnetic, or electric fields~\cite{asaad2020coherent}.
Combined with the deterministic incorporation pathways identified here for AlCl$_3$, this makes Al an intriguing dopant for single-acceptor applications while mitigating potential isotope variation.

\section*{Methods}
\label{sec:methods}

We use a Kinetic Monte Carlo model \cite{Bortz1975KMC,Gillespie1976KMC} as implemented in the \textsc{KMClib} package \cite{Leetmaa2014KMClib} to determine the probability of incorporation.
Our KMC model uses transition rates based on the Arrhenius equation $ \Gamma = A \exp{\Delta/k_{\rm B}T}$~\cite{arrhenius1889reaktionsgeschwindigkeit}, where $\Gamma$ is transition rate, $A$ is the attempt frequency, $\Delta$ is the reaction barrier found from our earlier DFT calculations, $k_{\rm B}$ is the Boltzmann constant, and $T$ is the temperature. 
We set all attempt frequencies $A$ to $10^{12}$ s$^{-1}$ as a reasonable order of magnitude estimate based on an analysis of attempt frequencies for the dissociation of phosphine on silicon~\cite{warschkow2016reaction}. 
We calculate the effusive flow rate of molecules landing on any particular silicon dimer as $\Phi_{effusion} =PA/\sqrt{2\pi m k_{\rm B}T}$, where $P$ is the pressure of the incoming precursor gas, $A$ is the area of impingement, taken here as a single silicon dimer, $m$ is the mass of the precursor gas.

Each KMC calculation is repeated 200 times with different random seeds, and the sample mean of the results is reported. 
We calculate error bars by assuming a binomial distribution of measured counts and using the standard error based on sample size.

\begin{center}
\begin{figure}
    \includegraphics[width=\columnwidth]{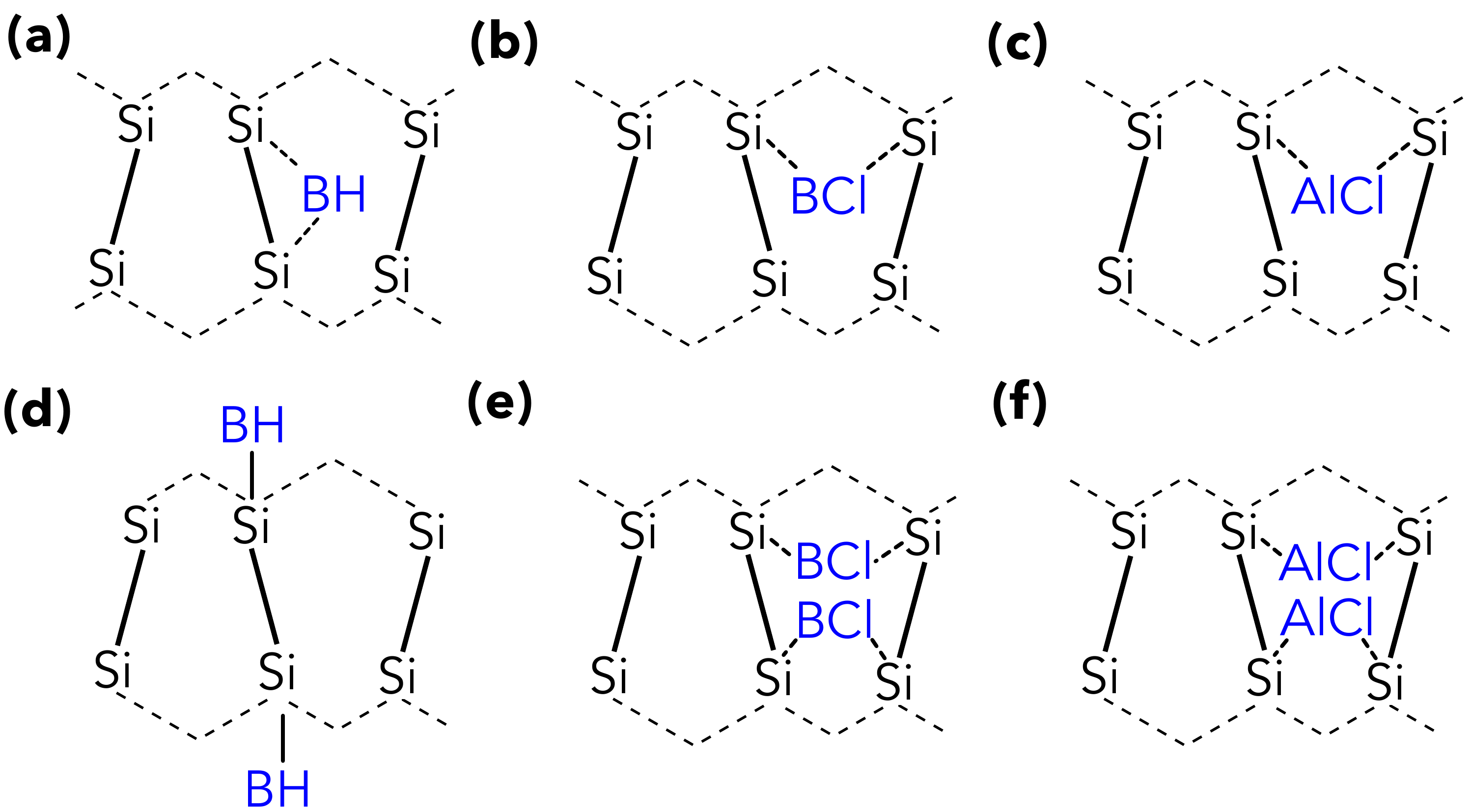}
    \caption{
    The endpoints accepted as incorporation events, or electrically inactive non-incorporation events by our kinetic model. 
    The bridging (a) BH, (b) BCl, or (c) AlCl configuration by itself is considered an incorporation for diborane, boron trichloride, and aluminum chloride respectively, due to the difficulty of desorption or reversing the dissociation reaction at these points.
    If two (d) BH, (e) BCl, or (f) AlCl form on the same dimer row(s), however, we assume that these dopants will dimerize and become electrically inactive and do not count these as incorporations within our model.
    }
    \label{fig:endpoints}
\end{figure}
\end{center}

Given the computationally prohibitive difficulty of modeling the full dissociation to  incorporation pathway, for all simulations we count a bridging BH, BCl, or AlCl on the silicon surface as an eventual incorporation, and all other fragments as non-incorporation events. 
This mirrors a similar treatment for phosphine from Warschkow~\emph{et al.}\cite{warschkow2016reaction}, for which Ivie, Campbell, Koepke~\emph{et al.} recently demonstrated agreement with incorporation rates inferred from STM scans for phosphine single donor incorporation within three dimer wide windows~\cite{ivie2021impact} that were consistent with prior analyses~\cite{Fuchsle2011thesis}.
In all cases, the adsorption energies of these bridging fragments are so low (typically below -2.0 eV) and the reverse reaction barriers so high (typically above 1.8 eV), that the chances of desorption from these bridging configurations are negligible, even at elevated anneal temperatures. 
Due to the likelihood of dimerized boron and aluminum being electrically inactive, we do not count any configurations where two BH, BCl, or AlCl exist on the same dimer row(s) as incorporations within our model.
The atomic configurations which we count as incorporations and disallow as electrically inactive dimerized incorporations are shown in Fig.~\ref{fig:endpoints}. 
Configurations are taken from Refs.~\cite{campbell2021model,dwyer2021impact,radue2021alcl3}.

\section*{Acknowledgements}
\label{sec:acknowledgements}
We gratefully acknowledge useful technical conversations with Jeff Ivie, Scott Schmucker, and Dave Wheeler 
We gratefully acknowledge Kenny Rudinger for fostering a group dynamic in which having a pun in the title of your paper seems like a good idea. 
This work was supported by the Laboratory Directed Research and Development program at Sandia National Laboratories under project 213017 (FAIR DEAL). 
Sandia National Laboratories is a multi-mission laboratory managed and operated by National Technology and Engineering Solutions of Sandia, LLC, a wholly owned subsidiary of Honeywell International, Inc., for DOE’s National Nuclear Security Administration under contract DE-NA0003525.
\section*{Author Contributions}

QC ran the computational simulations and analyzed results.
QC wrote the manuscript with significant input from ADB, REB, and SM.
ADB and SM conceived of and supervised the project. 

\section*{Data Availability}

All code for generating the data in this work is publicly available on github~\cite{kmccode}.

\section*{Competing Interests}

The authors declare no competing interests.

\bibliography{references.bib}


\end{document}